\newcommand{\lan}{\langle}
\newcommand{\ran}{\rangle}

\newcommand{\diracslash}[1]{#1\llap{/\kern2pt}}

\newcommand{\be}{\begin{equation}}
\newcommand{\ee}{\end{equation}}
\newcommand{\bea}{\begin{eqnarray}}\index{\footnote{}}
\newcommand{\eea}{\end{eqnarray}}
\newcommand{\ba}[1]{\begin{array}{#1}}
\newcommand{\ea}{\end{array}}

\documentclass[12pt]{iopart}

\usepackage{graphicx}
\usepackage{iopams}
\usepackage{revsymb}

\begin{document}
\setlength{\topmargin}{-0.05in}
\title{ Unitary quantum phase operators for bosons and fermions: A model study on quantum phases of 
interacting particles in a symmetric double-well potential}
\author{ Biswajit  Das$^1$, Bitan Ghosal$^{1*}$, Subhasish Dutta Gupta$^2$ and Bimalendu Deb$^{1,3}$ }
\address{$^1$ Department of Materials Science, $^3$ Raman Centre for Atomic, Molecular and Optical Sciences,
Indian Association for the Cultivation of Science,
 Jadavpur, Kolkata 700032. INDIA.
   $^2$ School of Physics, University of Hyderabad, Hyderabad 500 046. INDIA.}
\vspace{0.2in}
\hspace{0.95in} E-mail: msbd@iacs.res.in

\begin{abstract}
We introduce unitary quantum phase operators for material particles. 
 We  carry out a model study on quantum phases of interacting bosons in a symmetric 
   double-well potential in terms of unitary and commonly-used non-unitary 
    phase operators and compare the results for different number of bosons. We find that the results for unitary quantum phase 
   operators are significantly different from those for non-unitary ones especially in the case of low number of bosons.  
    We  introduce unitary operators corresponding to the quantum 
   phase-difference between two single-particle states of fermions. As an application of fermionic phase operators, we study 
   a simple model of a pair of interacting two-component fermions in a symmetric double-well potential. 
    We also investigate quantum phase and number 
 fluctuations to ascertain  number-phase uncertainty in terms of unitary phase operators. 
\end{abstract}

\maketitle

\section{Introduction }
The quantum phase  of interacting systems plays an important role in describing a variety of physical phenomena such as  phase transitions,
superfluid tunnelling, Josephson effects etc. 
With advancement in precision  interferometry with ultracold atoms 
\cite{prl:2004:Ketterle,nature:2007:bloch,prl:2007:Phillips} 
in confined geometries 
such as traps and atom chips, developing a proper understanding of the quantum phase properties of interacting
particles is of prime interest. As far as quantum phase measurement and its theoretical interpretations 
are concerned, there are unresolved issues which need to be addressed, in particular in the context of emerging field of atom optics.
For instance, a proper definition of `quantum phase' of electromagnetic fields had remained  
a hotly debated topic in theoretical quantum physics for a long time 
\cite{prsl:1927:Dirac,pl:1963:Louisell,p:1964:Susskind-Glogower,rmd:1968:Carrathers-Nieto,jpa:1986:Pegg-Barnett}.
Accurate determination of phase difference between two optical fields in the quantum domain 
remains an elusive task due to lack of  theoretical understanding of quantum phases. 
It is also known that such a difficulty exists  even in the case of semi-classical radiation theory when 
the field is weak and amplitude and phase fluctuations are correlated.
About two decades ago,  Mandel's group \cite{prl:1991:Mandel,pra:1992:Mandel,prl:1993:Mandel,pra:1994:Mandel} 
experimentally examined  two closely related but distinct measurement schemes for determining  
phase difference between two optical fields  in both semi-classical and quantum cases. 
They made use of sine and cosine of phase-difference operators as defined by Carruthers and Nieto \cite{rmd:1968:Carrathers-Nieto}.
Furthermore, to resolve the problems associated with  quantum phase measurement,
Noh  {\it et al.} \cite{prl:1991:Mandel,prl:1993:Mandel} introduced an operational definition of
quantum phase that requires  different phase operators for
different measurement schemes.

Quantum phase problems  for electromagnetic fields had been extensively studied during 90's.  However, to the best 
of our knowledge,  quantum phase problem in the context
 of matter waves  has not been  addressed so far. 
 Trapped ultracold atoms can be considered as an isolated interacting many-particle quantum system in the absence 
 of any appreciable trap-loss.
 It is then necessary to formulate quantum phase of  matter waves with fixed total number of particles. To introduce 
 quantum phase operators for material particles, one has to distinguish between bosonic and fermionic matter. Unitary quantum 
 phase operators for bosons are introduced following the  existing  quantum phase operator formalism of photons 
 which are bosons. Quantum phase operators for fermions are not known. It 
 is difficult  to define a unitary quantum phase operator for fermions
 by a simple extension of the existing quantum phase formalism of photons, because unlike photons,
 more than one fermion can not occupy a single quantum state.

 Here we introduce unitary and Hermitian  quantum phase operators corresponding to the phase-difference between 
 two single-particle quantum states in terms second quantised fermionic operators. A quantum state for fermions 
 can be either filled (by one fermion) or empty (vacuum state). 
 The unitarity of the operators is ensured by coupling the filled state with the vacuum.
 Therefore, quantum phase-difference
 between two fermionic modes 
 becomes well defined when single-particle quantum states of fermions are half filled. 
 To understand the underling features of quantum phases of interacting fermionic or bosonic  particles, 
 we resort to a simple microscopic model of  1D  symmetric double-well potential. 
 In case of two particles, 
 this model enables us to work with an analytical solution.
 In order to elucidate canonically conjugate nature of number-difference and phase-difference operators, 
we apply our formalism to study quantum fluctuations of both number- and phase-difference operators.
One can introduce two non-commuting operators corresponding 
to the cosine and sine of the phase-difference operators.   
 Both of them are canonically conjugate to the number-difference operators. 
These two phase operators plus the 
number-difference operator form a closed algebra. 
A unitary and hermition phase-difference operator and corresponding phase-difference state can be constructed using 
the cosine and sine  phase-difference operators.

 We find that for low number of bosons, the phase properties 
calculated with unitary phase-difference operators are significantly different from those calculated with non-unitary Carruther-Nieto 
phase-difference operators. However, for large number of bosons the results tend to converge.
Since unitarity of phase operators is ensured 
by coupling vacuum state with the highest number state in a finite dimensional Fock space, the effects 
of vacuum state on quantum 
phase properties is found to be quite substantial  in case of low number of bosons.  Since the operator corresponding to 
the number-difference between the two modes is canonically conjugate to the phase-difference operator, we also study the fluctuation
of number-difference to ascertain number-phase uncertainty and non-classical behaviour in quantum phase dynamics. 
In case  of fermions, we investigate 
fluctuations in phase-difference and number-difference of a pair of two-component interacting fermions.  

This paper is organised in the following way. In section $2$, we  introduce unitary quantum phase operators  of bosons 
as well as fermions and discuss in some detail the unitarity of quantum phase  operators. 
We describe hamiltonian dynamics of a few interacting bosons or fermions 
in a 1D symmetric double-well potential in section 3.
In section 4, we present and analyse numerical results on quantum phases,  number- and phase-fluctuations of different 
number of bosons in in the double-well potential. For fermions, we study numerically only the case of a pair of 
two-component fermions and compare the results with those of a pair of bosons.  We conclude this paper in section $5$.

\section{Quantum Phase Operators}

Here we present  operator formalism  for quantum phases of 
bosons and fermions.
For bosons, a proper quantum mechanical phase operator can be defined following that for quantised radiation fields. 
For
fermions, there exists no standard definition of a proper quantum mechanical phase operator. 
We here introduce a unitary  phase operator for fermions. 
Before we discuss our new formalism, let us  have a revisit into the history of quantum phase problem.

In classical 
mechanics, amplitude and phase are two canonically conjugate variables which appear 
in the expression of displacement  $ x_{\mu} (t) = 2 A_{\mu} \cos(\phi_{\mu} - \omega_{\mu} t)$  of  classical field of $\mu$-th mode.
This can be rewritten as 
\bea
x_{\mu} (t) = 2 A_{\mu} \cos(\phi_{\mu} - \omega_{\mu} t) = A_{\mu}  \left [ e^{i\phi_{\mu}}e^{-i \omega_{\mu} t} + 
 e^{-i\phi_{\mu}}e^{i \omega_{\mu} t} \right ] \label{xclassical} \eea
where $A_{\mu}$ and $\phi_{\mu}$ are the amplitude 
and phase of the $\mu$-th mode of the field and $\omega_{\mu}$ represents a harmonic frequency. In quantum mechanics, $x_{\mu}(t)$ is
replaced by the operator 
\bea 
\hat{x}_{\mu}(t) = x_{\mu 0} \left [  \hat{a}_{\mu} e^{-i\omega_{\mu} t} + \hat{a}_{\mu}^{\dagger} e^{i\omega_{\mu} t} \right ]
\label{xqm} \eea 
where  $\hat{a}_{\mu} (\hat{a}_{\mu}^{\dagger})$ represents the annihilation (creation) operator of a quanta of the quantised field in 
$\mu$-th mode. A comparison between the equations (\ref{xclassical}) and  (\ref{xqm}) suggests that in the classical limit 
$\hat{a}_{\mu} \rightarrow A_{\mu} e^{i \phi}$. Assuming that there exists a Hermitian phase operator $\hat{\phi}_{\mu}$ which 
is canonically conjugate to the number operator $\hat{N}_{\mu} = \hat{a}_{\mu}^{\dagger} a_{\mu}$, classical limit 
of such a phase operator may be attained by making use of the coherent state description of quantised fields. In classical mechanics, 
the azimuthal angle $\phi$ is given by 
\bea 
\phi = \tan^{-1}(y/x)
\eea 
and defines a modulo of $2\pi$. Now, defining $\phi$ to be continuous in $-\infty < \phi < \infty $, the angular momentum $L_z$ 
in three dimension is given by 
\bea 
L_z = x P_{y} - y P_x = -i \hbar \frac{\partial}{\partial \phi}
\eea
where $P_x$ and $P_y$ are $x$- and $y$-component, respectively, of the momentum $P$. $L_z$ and $\phi$ are conjugate variables 
and they satisfy 
\bea 
[\phi, L_z ] = i \hbar 
\eea 
$L_z$ is Hermitian in the space of periodic functions of period $2\pi$, but here $\phi$ is not periodic. Susskind and Glogower
\cite{p:1964:Susskind-Glogower} showed that  the non-periodicity of $\phi$ makes  $\phi$ non-hermitian.  

To define a phase operator in quantum mechanics is a delicate problem. The major difficulty in
defining a proper phase operator is its non-unitarity which stems from the fact that the number operator of a harmonic oscillator
has a lower bound in its eigenvalue spectrum. 
Dirac \cite{prsl:1927:Dirac} first postulated the existence of a Hermitian phase operator in his description of 
quantised electromagnetic fields. Susskind and Glogower \cite{p:1964:Susskind-Glogower} first showed that 
 Dirac's phase operator was not unitary and hence not Hermitian. Using Dirac's phase operator, if one tries to constructs 
a unitary operator $U$, then it turns out that, $UU^{\dagger}$ is an identity operator but  $U^{\dagger}U$ is not an 
identity operator. Therefore, $U$ is not unitary. Thus Susskind and Glogower \cite{p:1964:Susskind-Glogower} concluded 
that since $U$ is not unitary  there does not exist a Hermitian phase operator.

 Louisell \cite{pl:1963:Louisell} first introduced the periodic operator function in defining a phase
  variable conjugate to the angular momentum.
 Carrauthers and Nieto \cite{rmd:1968:Carrathers-Nieto} showed that one can define two Hermitian phase operators
 C and S corresponding to cosine and sine of the classical phase, respectively. However, these operators are non-unitary. 
 Using these operators, 
 they introduced  two-mode phase difference operators of a two-mode radiation field. 
 Explicitly, 
 the two-mode phase-difference operators are defined as
\bea
 \hat{C}_{12}^{\rm{CN}} = \hat{C}_1\hat{C}_2  + \hat{S}_1\hat{S}_2\nonumber\\
 \hat{S}_{12}^{\rm{CN}} = \hat{S}_1\hat{C}_2 - \hat{S}_2\hat{C}_1
\eea
where
\bea
 \hat{C}_i  = \frac{1}{2}((\hat{N}_i + 1)^{-\frac{1}{2}}\hat{a}_i + \hat{a}_i^{\dagger}(\hat{N}_i + 1)^{-\frac{1}{2}})
 \eea
 \bea
 \hat{S}_i  = \frac{1}{2i}(\hat{N}_i + 1)^{-\frac{1}{2}}\hat{a}_i - \hat{a}_i^{\dagger}(\hat{N}_i + 1)^{-\frac{1}{2}})
 \eea
 are the phase operators corresponding to sine and cosine, respectively, of the $i$-th mode.
In terms of creation (annihilation) operator  $\hat{a}_i^\dagger (\hat{a}_i)$  of the corresponding modes ($i=1,2$), 
phase difference operator can be written as
\begin{eqnarray}
 \hat{C}^{\rm{CN}}_{12} = \frac{1}{2}[(\hat{N}_1 + 1)^{-\frac{1}{2}} \hat{a}_1 \hat{a}_2^\dagger(\hat{N}_2 + 1)^{-\frac{1}{2}}
+\hat{a}_1^\dagger(\hat{N}_1 + 1)^{-\frac{1}{2}} (\hat{N}_2 + 1)^{-\frac{1}{2}} \hat{a}_2]
\eea
\bea
\hat{S}^{\rm{CN}}_{12} = \frac{1}{2i}[(\hat{N}_1 + 1)^{-\frac{1}{2}} \hat{a}_1 \hat{a}_2^\dagger(\hat{N}_2 + 1)^{-\frac{1}{2}}
-\hat{a}_1^\dagger(\hat{N}_1 + 1)^{-\frac{1}{2}} (\hat{N}_2 + 1)^{-\frac{1}{2}} \hat{a}_2]
\end{eqnarray}
The above cosine and sine phase-difference operators are non-unitary.  
Pegg and Burnett \cite{jpa:1986:Pegg-Barnett} first introduced a Hermitian and unitary phase operator.

In the description
of interference 
phenomena and interferometric experiments, we need to evaluate the phase-difference between two fields and not the absolute
phase of a field. It is therefore practical to seek   
a Hermitian  operator corresponding to phase-difference between two modes of a quantised field. 
By synthesizing the methods of Pegg-Burnett \cite{jpa:1986:Pegg-Barnett} and Carruthers-Nieto,
Deb {\it et al.} \cite{pra:1993:Deb} introduced Hermitian and unitary phase-difference operators of a
two-mode field with fixed number of total photons. It is done by  coupling  the vacuum state of one mode 
with the highest Fock state of the other in finite dimensional Fock space. 
The cosine and sine  phase-difference operators take form \cite{thesis:1996:Deb}
\begin{eqnarray}
\hat{C}_{12} =\hat{C}^{\rm{CN}}_{12}
+   \hat{C}_{12}^{(0)} \\
\hat{S}_{12} = \hat{S}^{\rm{CN}}_{12} + \hat{S}_{12}^{(0)}
\end{eqnarray}
where \bea \hat{C}_{12}^{(0)} = \frac{1}{2}\left [ |N, 0\rangle\langle 0, N|+|0, N\rangle\langle N, 0|\right] \eea
\bea \hat{S}_{12}^{(0)}  =
 \frac{1}{2i}\left[|N, 0\rangle\langle 0, N|-|0, N\rangle\langle N, 0| \right ] \eea
 describe the contributions from the vacuum states of the two modes.   $|N_1, N - N_1 \rangle$ represents a two-mode Fock state with $N_1$ and $N - N_1$ being the photon numbers in mode 1 and 2, respectively.
In case of quantised electromagnetic fields, the assumption of a fixed number of photons is made to circumvent the problem of 
non-unitarity. However, after all calculations are done one has to take the limit that the number of photons goes to infinity.

 \subsection{Unitary quantum phase operators for bosons} 
 
 Since it is possible to keep the total number of particles in a double-well fixed in the absence of any loss, 
 the assumption of a fixed total number of quanta (in this case, the total number of particles) is justified  
 and not just for a calculational advantage as in electromagnetic fields. We consider that the two modes 
 1 and 2 correspond to the second quantized matter wave fields of  
 the left  and right  well, respectively, of 1D double-well potential of equation (\ref{dwp}).
 We further assume that the energy is low and the inter-particle interaction is weak so that 
 a single boson remains in the lowest energy state with symmetric combination of the two harmonic oscillator 
 ground states. Let the operators $\hat{a}_i (\hat{a}_i^{\dagger})$ represent annihilation (creation)
 operator of a particle
 in left ($i=1$ or  right
well ($i=2$) harmonic oscillator ground state. 
Then the operators defined in equations (5) and (6) suffice to be the two-mode  phase-difference operators of bosons in a double-well potential corresponding to the cosine and sine, respectively, of the two-mode phase-difference. Thus we have $\hat{C}^{\rm{B}}_{12} \equiv \hat{C}_{12}$ and $\hat{S}^{\rm{B}}_{12} \equiv \hat{S}_{12}$ where the superscript $\rm{B}$ stands for boson. The difference of the number operators  or
the population imbalance between the two wells is $\hat{W}^{\rm{B}}=\hat{a}_1^\dagger\hat{a}_1-\hat{a}_2^\dagger\hat{a}_2$. The three operators $\hat{C}^{\rm{B}}_{12}$, $\hat{S}^{\rm{B}}_{12}$ and $\hat{W}^{\rm{B}}$ operators obey closed cyclic commutation algebra as follows
\bea
\left[[\hat{C}^{\rm{B}}_{12}, \hat{S}^{\rm{B}}_{12}], \hat{W}^{\rm{B}}\right] &+& \left[[\hat{S}^{\rm{B}}_{12}, \hat{W}^{\rm{B}}], \hat{C}^{\rm{B}}_{12}\right]
+\left[[\hat{W}^{\rm{B}}, \hat{C}^{\rm{B}}_{12}], \hat{S}^{\rm{B}}_{12}\right] = 0
\eea
The commutation algebra of the given operators are following
\begin{eqnarray}
\left[\hat{C}^{\rm{B}}_{12}, \hat{W}^{\rm{B}}\right] &=& 2i\left(\hat{S}^{\rm{B}}_{12} - (N+1)\hat{S}^{{(0)}}_{12}\right)\nonumber\\
\left[\hat{S}^{\rm{B}}_{12}, \hat{W}^{\rm{B}}\right] &=& -2i\left(\hat{C}^{\rm{B}}_{12} - (N+1)\hat{C}^{(0)}_{12}\right)\nonumber\\
\end{eqnarray}

One can define a unitary phase-difference  operator \cite{pio:1998:Perinova}
\bea
\hat{\beta}_{12} = \hat{C}^{\rm{B}}_{12} + i \hat{S}^{\rm{B}}_{12}. \label{qpop} \eea
The eigenstate of this operator, that is,  phase-difference  state can be constructed as the product of single-mode phase states of Pegg-Burnett \cite{jpa:1986:Pegg-Barnett} 
subject to the condition that total number of quanta in the two modes is a constant of motion. The procedure for deriving phase-difference state is described in references 
\cite{pra:1993:Deb,thesis:1996:Deb}. The important point to be noted here is that for low number of total bosons $N$, 
the effect of vacuum states such as $\mid 0,N \rangle$ and $\mid N,0\rangle $
on quantum phase-difference is quite  significant as would be discussed in section 4. Therefore, in case of low $N$, one has to use  unitary quantum phase-difference operators as defined in equation 
(\ref{qpop}) for accurate  measurement of quantum phases. The phase-difference operators of Carruthers and Nieto will approach unitarity  in the limit $N \rightarrow \infty $. 
 In a recent theoretical paper by Sarma and Zhou \cite{pra:2010:Das-Sarma}, 
 an operator similar to that of Carrruthers and Nieto has been implicitly used for studying phase dynamics
 of a Bose-Einstein condensate (BEC) in a  double-well. 
 It is worthwhile to mention that while in case of BEC, probably a non-unitary phase operator such as used in  
\cite{pra:2010:Das-Sarma} can suffice for phase measurement for all practical purpose, 
a few bosons in a double well necessarily require unitary phase difference operators for
 high precision phase measurement, in particular for the purpose quantum information processing with a few bosons in a double-well.

\subsection{Unitary quantum phase operators for  fermions}

In the previous subsection, we have introduced 
phase operators for  massive bosons  with an 
 analogy with the phase operators of electromagnetic fields of  massless photons. This analogy has been 
possible because of bosonic symmetry in both the cases. Bosonic symmetry allows a large number 
of bosons to occupy a  single-particle quantum state (or mode). 
For fermions, obviously such an analogy can not be drawn. A single-particle quantum state defined by a set of quantum numbers 
including spin magnetic quantum number can not be occupied by more than one 
fermion due to  Pauli's exclusion principle.  

\begin{figure}
\includegraphics[height=3.25in, width=\columnwidth]{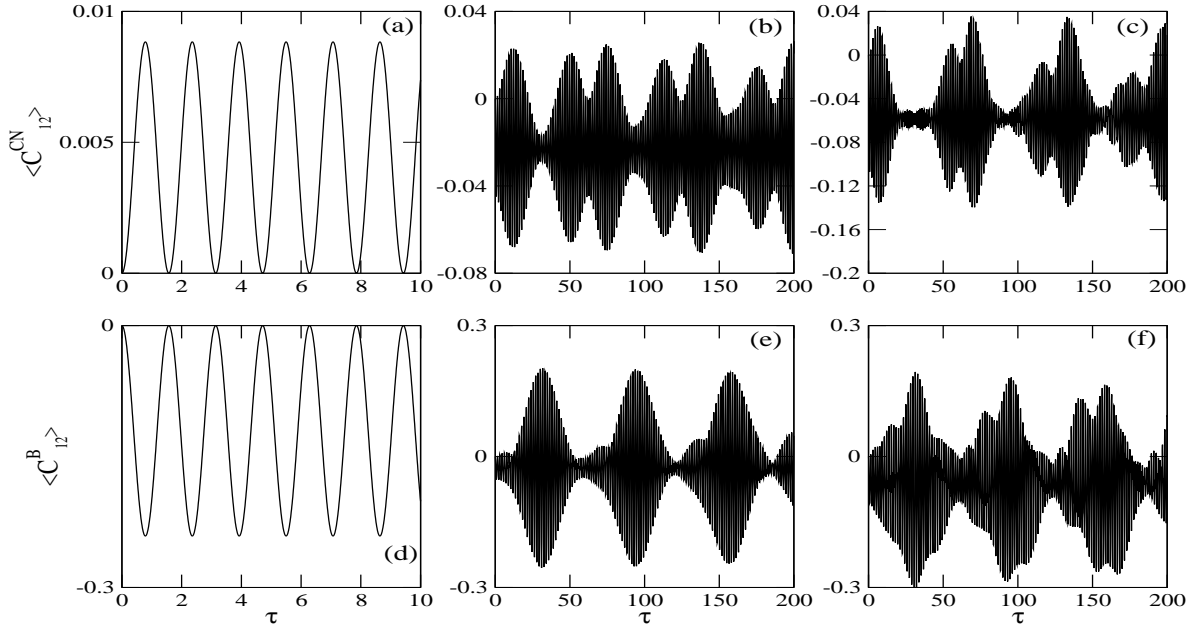}
\caption{ The top and bottom panels show the variation of average of the cosine phase-difference operator 
$\langle C^{\rm{CN}}_{12}\rangle$ and $\langle C^{\rm{B}}_{12}\rangle$, respectively,  
as a function of $\tau$ for $\bar U=0.05$. 
From left to right, subplots correspond to
different number of bosons $N=2$ (a,d), $N=5$ (b,e) and $N=10$ (c,f) respectively.}
\label{fig1}
\end{figure}
 
 \begin{figure}
\includegraphics[height=3.25in, width=\columnwidth]{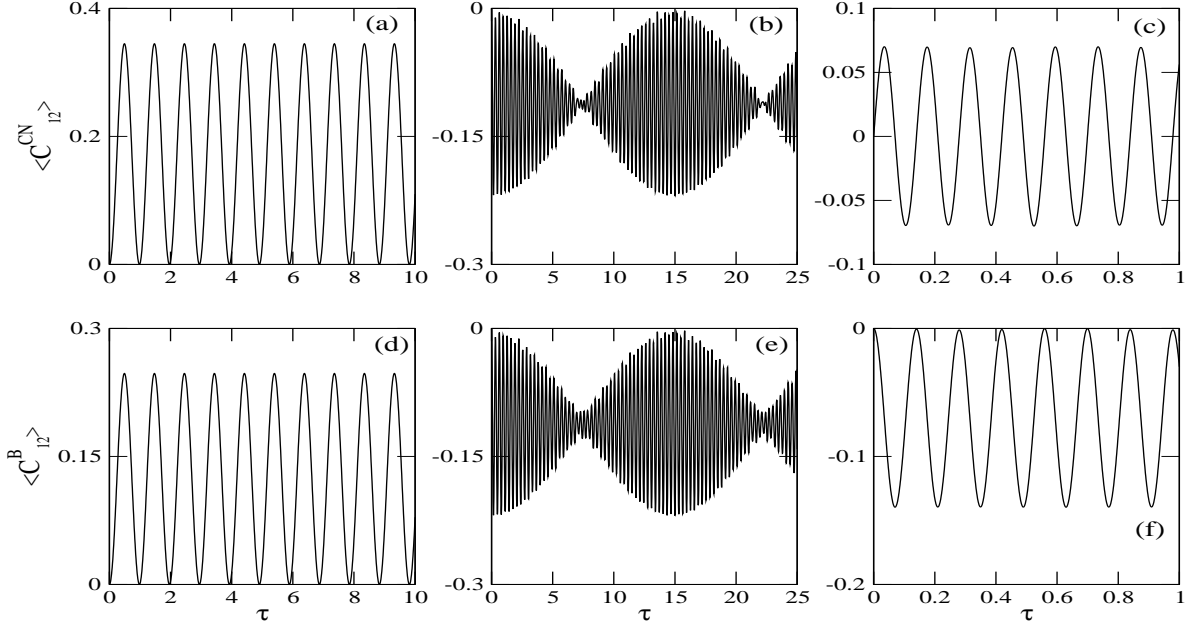}
\caption{Same as in figure 1 but  for $\bar U=5.0$.
The subplots correspond to different number of bosons $N=2$ (a,d), $N=5$ (b,e) and $N=10$ (c,f) 
respectively.}
\label{fig2}
\end{figure}

Let us now discuss how we can introduce operators corresponding to the phase-difference between 
two single-particle fermionic quantum states or modes.  If we construct
a phase-difference operator between two fermionic modes following the method of Carruthers and Nieto, 
obviously such an operator will deviate largely from the unitarity since a fermionic mode can be occupied more than 
a single fermion. Let us write two  hermition operators corresponding to the cosine and sine of phase-difference 
between two fermionic modes along the lines of Carrathers-Nieto phase formalism and check the
usefulness of such operators. Two fermionic modes can be characterised by the the principal quantum numbers $n$ and $n'$ 
of low lying harmonic oscillator states on the left and right well, respectively,
and two spin degrees-of-freedom  $\sigma$ 
and $\sigma'$ on the left and right well, respectively. For simplicity, we assume that $n = n'$ and suppress $n$-index. 
We then have the two modes 
$m \equiv l, 
\sigma  $ 
and $m' \equiv r, \sigma' $. The cosine and sine phase-difference operators can be given by 
\bea
 \hat{C}_{m m'} &=&
 \frac{1}{2}\left[(\hat{N}_{m} + 1)^{-\frac{1}{2}}\hat{a}_{m}\hat{a}_{m'}^\dagger(\hat{N}_{m'} + 1)^{-\frac{1}{2}}
 \right. \nonumber \\
 &+& \left. \hat{a}_{m}^\dagger(\hat{N}_{m} + 1)^{-\frac{1}{2}}(\hat{N}_{m'} + 1)^{-\frac{1}{2}}\hat{a}_{m'}\right]
 \label{fccn}
\eea
 \bea
  \hat{S}_{m m'}
 &=&\frac{1}{2i}\left[(\hat{N}_{m} + 1)^{-\frac{1}{2}}\hat{a}_{m}\hat{a}_{m'}^\dagger(\hat{N}_{m'} + 1)^{-\frac{1}{2}}
 \right. \nonumber \\
 &-& \left.  \hat{a}_{m}^\dagger(\hat{N}_{m} + 1)^{-\frac{1}{2}}(\hat{N}_{m'} + 1)^{-\frac{1}{2}}\hat{a}_{m'}\right]
 \label{fscn}
 \eea
Here the operators $a_m$ satisfy the anti-commutator algebra 
\bea 
\left \{ a_m, a_{m''}^{\dagger} \right \} = a_m a_{m''}^{\dagger} + a_{m''}^{\dagger} a_m = \delta_{m m''} \equiv 
\delta_{s s''} \delta_{\beta \beta''} \delta_{\sigma \sigma''} 
\eea
\bea 
\left \{ a_m, a_{m''}\right \} = a_m a_{m''} + a_{m''} a_m  = 0
\eea
\bea 
\left \{ a_m^{\dagger}, a_{m''}^{\dagger} \right \} = a_m^{\dagger} a_{m''}^{\dagger} + a_{m''}^{\dagger} a_m^{\dagger} = 0
\eea
and $\hat{N}_m = a_m^{\dagger} a_m$ is the number operator of the $m$-th mode. The total number of fermions 
$N = \sum_{m} \hat{N}_m$ is assumed to be a constant.

To construct an eigenstate of   $\hat{C}_{m m'}$ or  $\hat{S}_{m m'}$, 
we consider  many-fermion basis states  in all possible configurations of fermion distribution 
in all available low-energy single-particle states. For instance, let us denote such  a basis state in the form      
 $\mid 0  1  \rangle_i \equiv 
\mid \{0\}_m, \{1\}_{m'} \rangle \otimes \mid \{(N-1)\} \rangle_i $  in $i$-th configuration in which 
 $\mid \{0\}_m, \{1\}_{m'} \rangle$ implies 0 fermion in $m$-th mode and 1 fermion in $m'$-th. The rest 
of the fermions ($N-1$) are distributed over all the modes except the two modes $m$ and $m'$. The ket 
$\mid \{(N-1)\} \rangle_i $ represents a configuration  of ($N-1$) fermions distributed over ($N_s -2$) modes.  A general many-fermion 
state can be written as 
\bea 
\mid \psi \rangle &=& \sum_{i } d_{00}^{(i)} \mid 0 0 \rangle_i + 
\sum_{j}  d_{01}^{(j)} \mid 0 1\rangle_j + 
\sum_{k}  d_{10}^{(k)} \mid 1 0\rangle_k 
+ \sum_{l}  d_{11}^{(l)} \mid 1 1\rangle_l 
\label{gs}
\eea
where $ \mid n  n'  \rangle_i \equiv 
\mid \{n\}_m, \{n'\}_{m'} \rangle \otimes \mid \{(N-n - n')\} \rangle_i $.  The summation over the index 
$i$ implies  sum over all the configurations in which $m$ and $m'$ can be empty. Similarly summation over 
$j$, $k$ and $l$  imply  sums over configurations of the types  $\mid 0 1\rangle_j$, $\mid 1 0\rangle_k$ 
and  $\mid 1 1\rangle_l$, respectively. From eigenvalue equation 
$ \hat{C}_{m m'} \mid \psi \rangle = C \mid \psi \rangle $ we infer that $d_{00}^{(i)} = 0$ 
for all $i$, $d_{11}^{(l)} = 0$ for all $l$  while $d_{01}^{(j)}$ and  $d_{10}^{(k)}$ are in general 
nonzero. Now, using the operators of equations (\ref{fccn}) and (\ref{fscn}), one can construct an operator 
$\hat{U}_{m m'} = \hat{C}^{\rm{F}}_{m m'} + i \hat{S}^{\rm{F}}_{m m'} = 
(\hat{N}_{m} + 1)^{-\frac{1}{2}}\hat{a}_{m}\hat{a}_{m'}^\dagger(\hat{N}_{m'} + 1)^{-\frac{1}{2}}$. 
Assuming a phase-difference state of the form $\mid \psi_{m m'} \rangle = \sum_{i}  d_{01}^{(i)} \mid 0 1\rangle_i + 
\sum_{k}  d_{10}^{(j)} \mid 1 0\rangle_j $ with $\sum_{i} | d_{01}^{(i)} |^2 + \sum_{j} |d_{10}^{(j)}|^2 = 1$,   
we find  $\langle \psi_{m m'} \mid \hat{U}_{m m'} \hat{U}_{m m'}^{\dagger} \mid \psi_{m m'} \rangle = \sum_i | d_{01}^{(i)} |^2 < 1 $ 
and  $\langle \psi_{m m'} \mid \hat{U}_{m m'}^{\dagger} \hat{U}_{m m'}   \mid \psi_{m m'} \rangle = \sum_j | d_{10}^{(j)} |^2 < 1 $. 
Thus we notice that  $\hat{U}_{m m'}$ does not satisfy unitarity. Unlike large number of bosons an unitarity limit for 
these fermionic operators does not exist and therefore it can be concluded that for fermions Carruther-Nieto type 
phase operators for fermions do not exist. 

Let us now investigate whether it is possible to have a unitary fermionic phase-difference operator by adding new terms 
as done in bosonic case. This amounts to  coupling zero-fermion 
state (vacuum) with one-fermion state  (highest number state in case of fermion). By doing so, we have new cosine and sine phase-difference operators for fermions in 
form 
\bea
 \hat{C}_{m m'}^{\rm{F}} &=&
 \frac{1}{2}\left[(\hat{N}_{m} + 1)^{-\frac{1}{2}}\hat{a}_{m}\hat{a}_{m'}^\dagger(\hat{N}_{m'} + 1)^{-\frac{1}{2}}
 + \hat{a}_{m}^\dagger(\hat{N}_{m} + 1)^{-\frac{1}{2}} \right. \nonumber \\
 &\times & \left. (\hat{N}_{m'} + 1)^{-\frac{1}{2}}\hat{a}_{m'}\right] + 
 \frac{1}{2} \sum_{i j} \left [ \mid 1 0 \rangle_{i}  \hspace{0.01cm} _{j}\langle 01 \mid + 
 \mid 0 1 \rangle_{j}  \hspace{0.01cm} _{i}\langle 1 0 \mid \right ] 
 \label{nfccn}
\eea
 \bea
  \hat{S}_{m m'}^{\rm{F}}
 &=&\frac{1}{2i}\left[(\hat{N}_{m} + 1)^{-\frac{1}{2}}\hat{a}_{m}\hat{a}_{m'}^\dagger(\hat{N}_{m'} + 1)^{-\frac{1}{2}}
 -  \hat{a}_{m}^\dagger(\hat{N}_{m} + 1)^{-\frac{1}{2}} \right. \nonumber \\
 &\times& \left. (\hat{N}_{m'} + 1)^{-\frac{1}{2}}\hat{a}_{m'}\right]
 + \frac{1}{2 i} \sum_{i j} \left [ \mid 1 0 \rangle_{i}  \hspace{0.01cm} _{j}\langle 01 \mid -  
 \mid 0 1 \rangle_{j}  \hspace{0.01cm} _{i}\langle 1 0 \mid \right ] 
 \label{nfscn}
 \eea
Now, constructing an operators $\hat{U}  = \hat{C}_{m m'} + i \hat{S}_{m m'} $, it is easy to verify that 
$\langle \psi_{m m'} \mid \hat{U}
\hat{U}^{\dagger} \mid \psi_{m m'} \rangle =   \langle \psi_{m m'} \mid \hat{U}^{\dagger} 
\hat{U} \mid \psi_{m m'} \rangle = \sum_{i} |d_{01}^{(i)} |^2 + \sum_j | d_{10}^{(j)}|^2 = 1 $. Thus, 
these phase operators are unitarity. It can be further verified that the operators $\hat{C}_{m m'}^{\rm{F}}$, 
$\hat{S}_{m m'}^{\rm{F}}$ and the number-difference operator $\hat{W}_{m m'}^{\rm{F}}$ satisfy a closed commutator 
algebra 
\begin{eqnarray}
\left[[\hat{C}^{\rm{F}}_{m m'}, \hat{S}^{\rm{F}}_{m m'}], \hat{W}_{m m'}^{\rm{F}}\right] &+& \left[[\hat{S}^{\rm{F}}_{m m'}, 
\hat{W}_{m m'}^{\rm{F}}], \hat{C}^{\rm{F}}_{m m'}\right] \nonumber \\
&+& \left[[\hat{W}_{m m'}^{\rm{F}}, \hat{C}^{\rm{F}}_{m m'}], \hat{S}^{\rm{F}}_{m m'}\right] = 0
\end{eqnarray}

Before concluding this subsection it is worth mentioning that 
the unitary quantum phase-difference operators of equations (\ref{nfccn}) and (\ref{nfscn}) are most general 
in the sense that they apply to any fermionic system.

\section{Interacting particles in 1D double-well potential}

As an application of our unitary phase operators,  we consider a model of a few interacting bosons or fermions 
at low energy in a 1D symmetric double-well potential.  
The interaction is assumed to be of zero-ranged contact type.

\begin{figure}
\includegraphics[height=3.25in, width=\columnwidth]{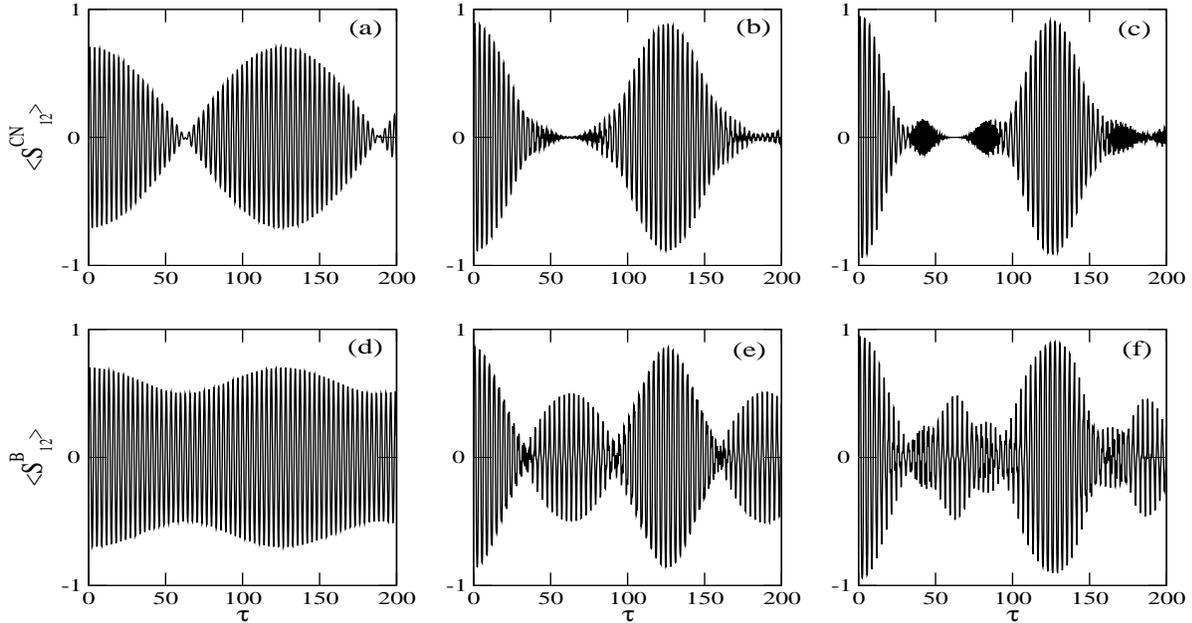}
\caption{ The top and bottom panels show
the variation of average of the sine phase-difference operator
$\langle S^{\rm{CN}}_{12}\rangle$ and $\langle S^{\rm{B}}_{12}\rangle$ as a function of
$\tau$ for $\bar U=0.05$. From left to right, subplots correspond to different number of bosons $N=2$ (a,d), $N=5$ (b,e) and $N=10$ (c,f),
respectively.}
\label{fig3}
\end{figure}

\begin{figure}
\includegraphics[height=3.25in, width=\columnwidth]{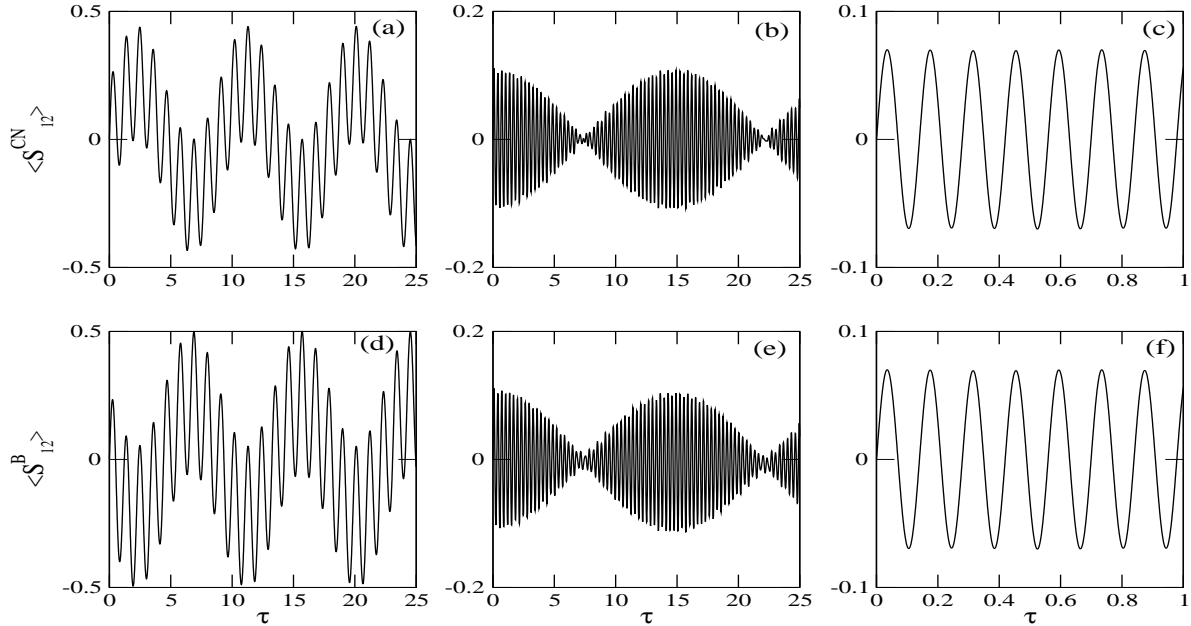}
\caption{Same as in figure 3 but  for $\bar U=5.0$.
The subplots correspond to different number of bosons $N=2$ (a,d), $N=5$ (b,e) and $N=10$ (c,f), respectively. }
\label{fig4}
\end{figure}

\subsection{A model double-well potential}
A model double-well potential can be written in different forms in 1D or 2D or 3D. 
To study quantum phase operators for massive particles,  
 we consider, for simplicity, a model of  one-dimensional symmetric double-well potential of the form
 \bea 
 V(x) = \frac{1}{2} \lambda^2 (x^2 - q^2)^2 \label{dwp} 
 \eea particles
 which has two minima at $x = \pm q$ and $\lambda$ is a parameter that 
 determines the barrier height $V_0$ between the two wells given by $V_0 = \lambda^2 q^4/2$. Expanding 
 this potential in Taylor series around $x_{\pm} = \pm q$, one finds that the leading order terms are quadratic 
 in $(x - x_{\pm})$ and therefore small oscillations around the positions $x_{\pm}$ are simple harmonic in the leading 
 order approximation. At the bottom of the wells around the positions $x_{\pm}$, the quantized motion 
 of a single particle may be approximated as that of simple harmonic oscillator. Let us call the well 
 at $x_{-} = -q$ as the left ($l$) or first (1) well and that at $x_{-} = q$ as the right ($r$) or second (2) one. In 
 the second quantised notation, annihilation (creation) of a single-particle harmonic oscillator ground state 
 at left (right) well can be described in terms of operators $a_{l(r)}$ ($ a_{l(r)}^{\dagger}$) or equivalently 
 $a_{1(2)}$ ($ a_{1(2)}^{\dagger}$). These harmonic oscillator states of individual wells are perturbative states when 
 the tunnelling  between the wells is neglected. In the presence of tunnelling, the two lowest eigenstates
 of a particle  in a symmetric double well 
 potential of the form (\ref{dwp}) are  symmetric and antisymmetric combination of the 
 two harmonic oscillator ground states which are not degenerate. 

Recently, double-well potentials have become  important in  research with ultracold atoms in  
traps and optical lattices, particularly in the context of few-body quantum dynamics \cite{schmelcher1,schmelcher2}, 
quantum tunnelling \cite{prl:2005:Albeiz,pra:1997:Milburn, prl:2008:Kierig}, 
Josephson oscillations \cite{prl:2005:Albeiz}, nonlinear self trapping \cite{prl:2005:Anker,prl:1986:Javanainen}
correlated pair tunnelling \cite{nature:2007:bloch,prl:2011:bloch,schmelcher2},  
number squeezing \cite{prl:2007:Phillips}, quantum magnetism \cite{science:2008:Trotzky} etc. 
  Theoretical studies with interacting atoms in a double-well have demonstrated  entanglement in atomic hyperfine spin and 
phase variables \cite{pra:2008:bloch}, an interplay between interaction and disorder in a BEC \cite{pra:2010:Das-Sarma}, operation of  
a quantum gate  \cite{ajp:2011:Foot-Shotter} and so on. With increasing use of double-wells in cold atom research, 
double-well optical lattice
 \cite{jpb:2006:Anderlini,pra:2006:Sebby,prl:2008:Vladimir,prl:2007:Lee} is emerging as 
an important tool for studying correlation effects in cold atoms \cite{nature:2007:Anderlini}.
Thus, addressing quantum phase
problems in a toy model such as interacting particles in a symmetric double-well potential is relevant and timely 
in the context of current cold atom research. There are several  advantages of this model. 
This model is  generalizable for optical lattice where one can  study 
  inter-site quantum phase fluctuations and their effects on Bose-Hubbard physics 
  \cite{prb:1989:Fisher,epl:1993:Ramakrishnan,nature:2002:Greiner,prl:1999:Jaksch} and 
  superfluid-Mott insulator quantum phase transition \cite{nature:2002:Greiner}.

  \begin{figure}
\includegraphics[height=3.25in, width=\columnwidth]{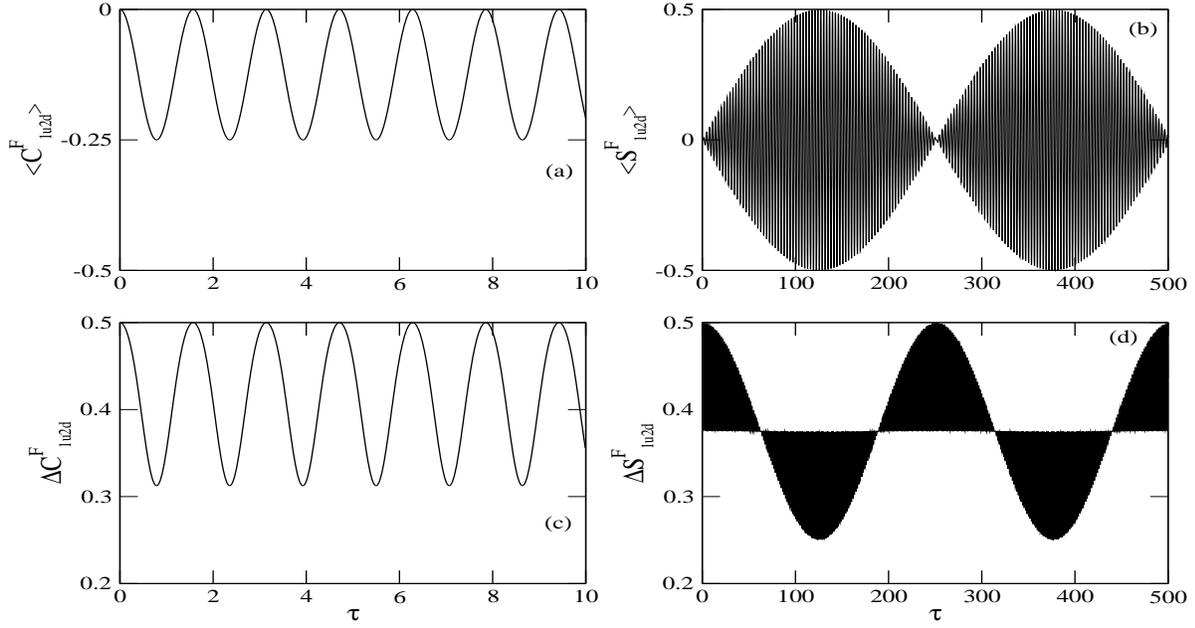}
\caption{The subplot (a) shows the average of the fermionic cosine phase-difference operator $<C_{1u2d}>$ 
between the spin $\uparrow$ ($u$) in the left 
well (1) and the spin $\downarrow$ ($d$) in the right well (2) as a function of dimensionless time $\tau$ for 
$\bar{U} = 0.05$. The subplot 
(b) shows the average of sine phase-difference operator between the same spin state. The subplots 
(c) and (d) display the fluctuations of the corresponding cosine and sine phase-difference operators.}
\label{fig5}
\end{figure}

\begin{figure}
\includegraphics[height=3.25in, width=\columnwidth]{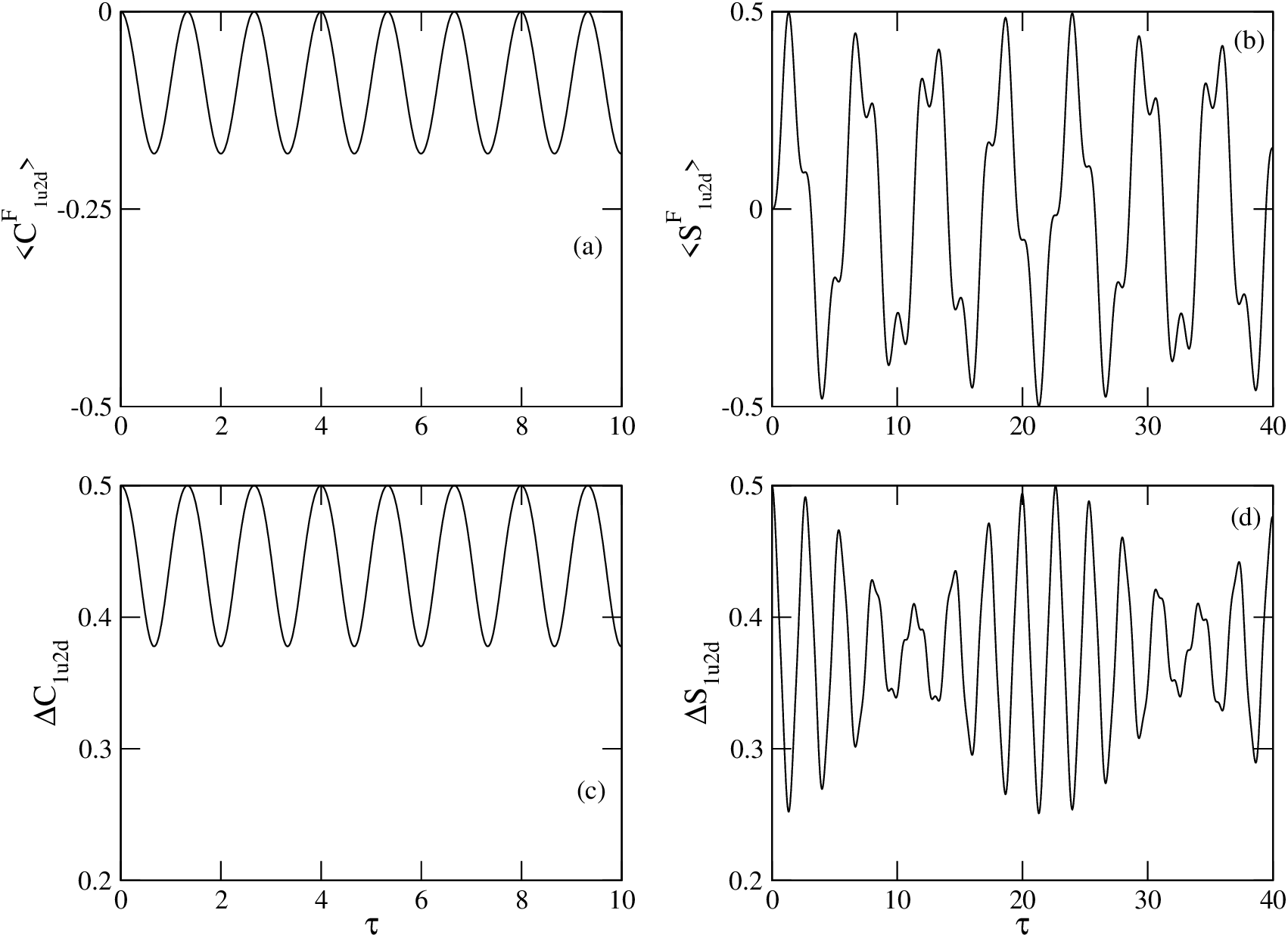}
\caption{Same as in figure 5 but for $\bar{U} = 5$. }
\label{fig6}
\end{figure}

\subsection{Bosons}
The Hamiltonian of a system of $N$ interacting bosons occupying two weakly coupled lowest states of a symmetric double well potential is given by
\begin{equation}
\hat{H}=-\hbar J(\hat{a}_l^\dagger \hat{a}_r +\hat{a}_r^\dagger \hat{a}_l)+\frac{\hbar U}{2}({\hat{a}_l^{\dagger 2}} \hat{a}_l^2+{\hat{a}_r^{\dagger 2}}\hat{a}_r^2)
\end{equation}
where $\hat{a}_{l,r}(\hat{a}_{l,r}\dagger)$ are the bosonic particle annihilation(creation) operators for the two sites $l$ (left) and $r$ (right) of the double well, $J>0$ accounts for the hopping or tunnelling between the two sites and $U$ is the on-site interaction.
The wave function $|\psi(t)\rangle$ in the basis of Fock states with fixed total particle number $N$ can be written as
\begin{equation}
|\psi(t)\rangle = \sum_{n_l=0}^{N} c_{n_l}(t) |n_l,N-n_l\rangle
\end{equation}
Where $c_{n_l}(t)$ is the probability amplitude to find $n_l$ particles in the left well and $(N-n_l)$ particles in the right well and $|n_l,N-n_l\rangle$ denote Fock state with $n_l$ particle at the left site and $N-n_l$ particle at the right site. From schr\"{o}dinger equation , we obtain
\begin{equation}
 i\frac{dc_l(t)}{dt} =-(\kappa_{l}c_{l+1}+\kappa_{l-1}c_{l-1})+V_{l}c_{l}
\end{equation}
where $\kappa_l = J\sqrt{(l+1)(N-l)}$,
$V_l = \frac{U}{2}[l^2+(N-l)^2-N]$
and the normalization condition is $\sum\limits_{l=0}^{N} |c_l(t)|^2=1$.
The bosonic model we  use here is similar to the one studied  by Longhi \cite{jpb:2011:Longhi} who has shown that 
the average dynamical behaviour of a pair of  hard-core bosons in a symmetric double-well has a classical counterpart
in the transport of 
electromagnetic waves through wave-guide arrays.  However, such an analogy can not be drawn for fermions in general.

In the special case of $N=2$, the solution is simple and  analytically tractable. For instance,
let us consider the initial condition $c_0(0)=1$ and $c_1(0)=c_2(0)=0$ i.e. both particles are initially in the right well. One then obtains
\begin{eqnarray}
c_1(t) &=&  C\exp\left[-i(\frac{\bar{U}}{2} - \Omega_{\rm{B}})\tau\right] + D\exp\left[-i(\frac{\bar{U}}{2} + \Omega_{\rm{B}})\tau\right]\nonumber\\
 c_2(t) &=& c_2(0) + i\sqrt{2}\left[C\frac{\exp[-i(\frac{\bar{U}}{2} - \Omega_{\rm{B}})\tau] - 1}{-i(\frac{\bar{U}}{2}-\Omega_{\rm{B}})}
 + D\frac{\exp[-i(\frac{\bar{U}}{2} + \Omega_{\rm{B}})\tau] - 1}{-i(\frac{\bar{U}}{2}+\Omega_{\rm{B}})}\right]\nonumber\\
 c_0(t) &=& c_0(0) + i\sqrt{2}\left[C \frac{\exp[-i(\frac{\bar{U}}{2} - \Omega_{\rm{B}})\tau] - 1}{-i(\frac{\bar{U}}{2}-\Omega_{\rm{B}})}
 + D \frac{\exp[-i(\frac{\bar{U}}{2} + \Omega_{\rm{B}})\tau] - 1}{-i(\frac{U}{2}+\Omega_B)}\right]\nonumber\\
\end{eqnarray}
where $\bar{U}=U/J$, $\tau=Jt$, $\Omega_{\rm{B}} = \sqrt{4 + (\bar{U}/2)^2}$, $ C = \frac{(\Omega_{\rm{B}} - \frac{\bar{U}}{2}) c_1(0) + \sqrt{2}(c_2(0)+c_0(0))}{2 \Omega_{\rm{B}}}$ and
 $ D = \frac{(\Omega_{\rm{B}} + \frac{\bar{U}}{2}) c_1(0) - \sqrt{2}(c_2(0)+c_0(0))}{2 \Omega_{\rm{B}}}$. Assuming initial condition $c_0(0)=1$ and $c_1(0)=c_2(0)=0$,  i.e. both bosons are initially in the right well, we get
$ C = \frac{1}{\sqrt{2} \Omega_{\rm{B}}}$ and  $ D = - \frac{1}{\sqrt{2} \Omega_{\rm{B}}}$.
Note that these solutions are the same as in \cite{jpb:2011:Longhi}.

\subsection{Fermions}

For many fermions in a symmetric double-well potential, it is essential  to take 
into account a large number of single-particle states even at ultra low energy. This means 
that one has to consider excited states of harmonic oscillators, but then harmonic approximation 
of the double-well potential of (\ref{dwp}) may break down. 
If one considers a few interacting two-component fermions in first two eigenstates of the 
harmonic oscillators around $x = \pm q$ of equation  (\ref{dwp}), one has  to take into account two on-site 
interaction parameters - one between fermions with same spin (triplet) states and the other between
dissimilar spin (singlet) states. Also, one has to consider two tunnelling terms - one between harmonic oscillator
ground states and another between excited states. Thus a few fermion dynamics in a symmetric double-well 
potential becomes quite complicated. As an example of applications of the unitary fermion quantum phase
operators, for simplicity, we assume that the kinetic energy of 
individual fermions is low enough so that they can occupy 
the low lying states of the two symmetric harmonic oscillators in the limit 
$\lambda \rightarrow \infty$ of the symmetric double-well potential of (\ref{dwp}). It is assumed that 
the interaction between 
fermions is due to a contact potential.   The fermion-fermion on-site interaction $U$ is assumed to be much 
smaller than  the harmonic frequency.  
We then  assume that the lowest harmonic oscillator states ($n=0, n'=0$) of the two wells
are occupied with all other higher states being empty. Under these conditions, for spin-polarised fermions, there are 
2 possible single-particle 
quantum states, and for two-component fermions there are 4 single-particle quantum states 
corresponding to  the harmonic ground states of the two sites $s= l,r$ where $s$ stands for site index.   Now, 
if all these available  low energy 
levels are filled up, then at low energy tunnelling dynamics of fermions may be suppressed due to Pauli blocking.  
For our studies we assume that quantum states are half-filled. 

\begin{figure}
\includegraphics[height=3.25in, width=\columnwidth]{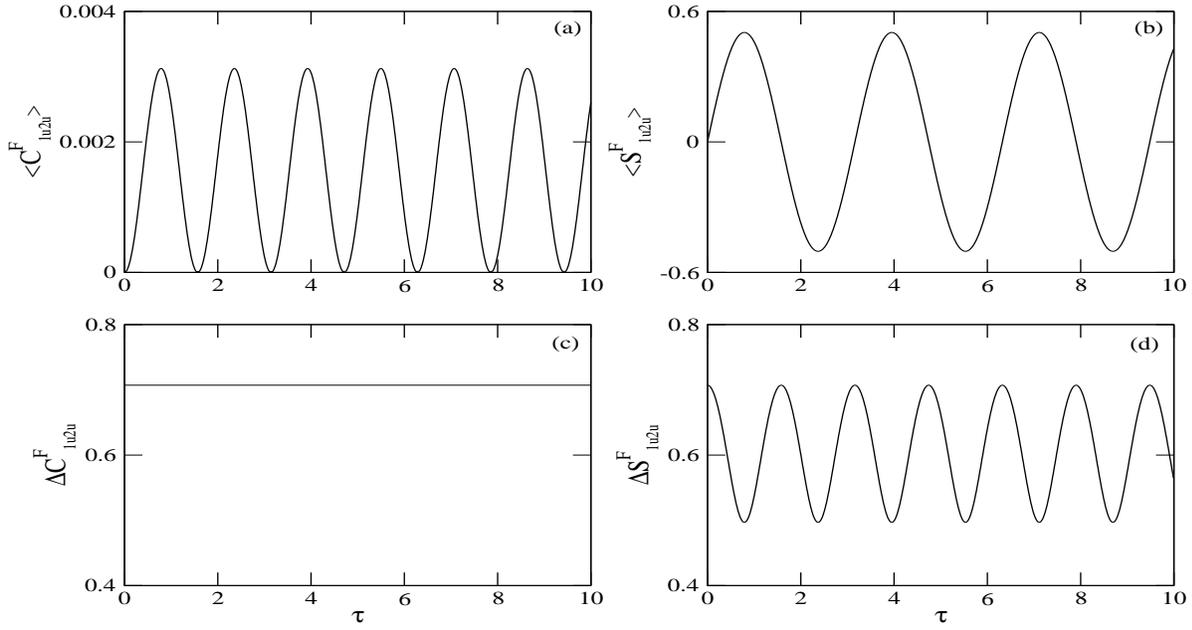}
\caption{Same as in figure 5 but for the phase-difference operators between the spin $ 1 \uparrow$ ($1u$) 
and $ 2 \uparrow$ ($2u$). }
\label{fig7}
\end{figure}

\begin{figure}
\includegraphics[height=3.25in, width=\columnwidth]{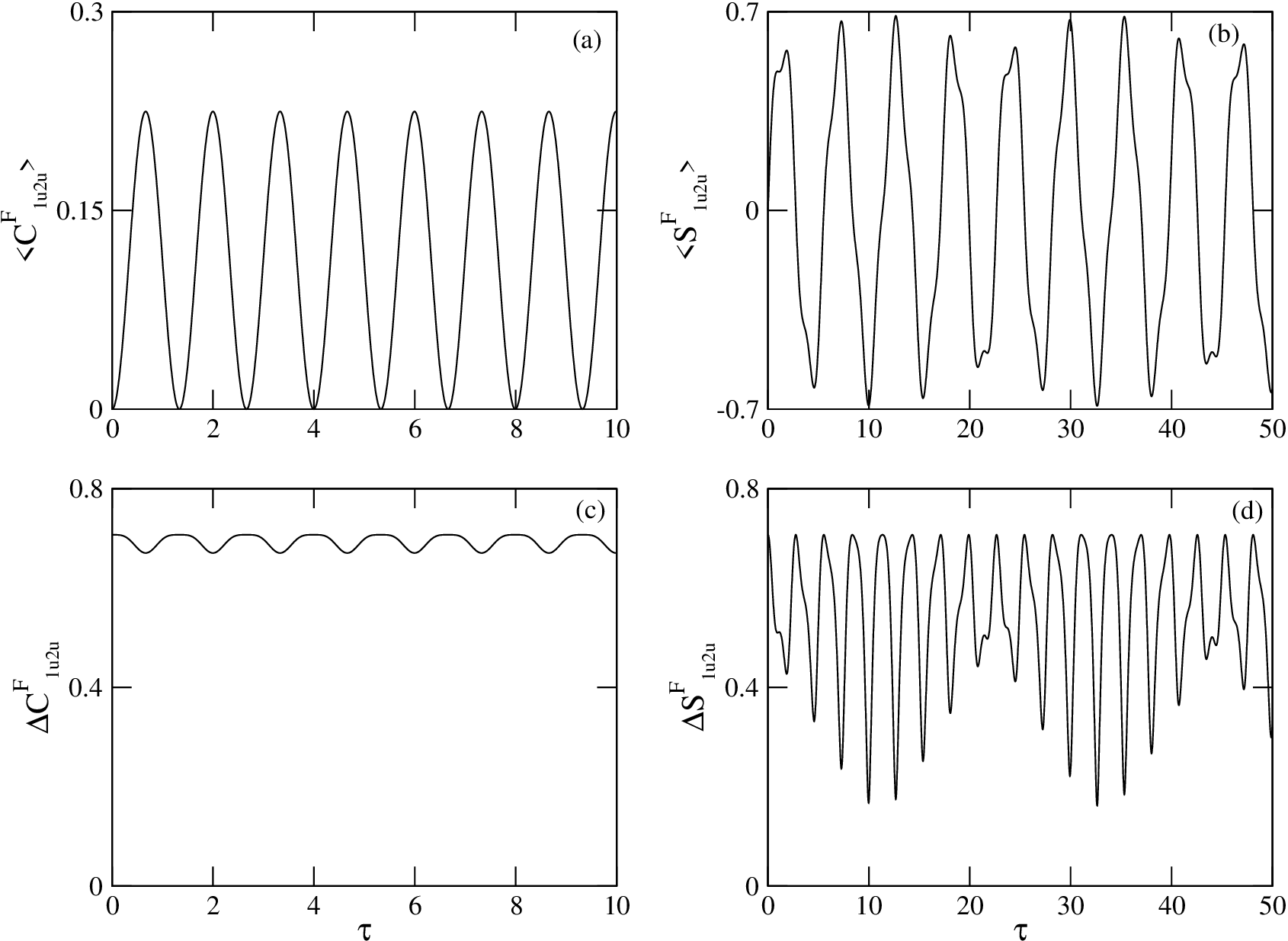}
\caption{Same as in figure 7 but for $\bar{U} = 5$ }
\label{fig8}
\end{figure}

Here we deal with only two-component fermions, one with spin up and another with spin down.
The Hamiltonian of the system of a pair of two-component fermions  is given by
\begin{eqnarray}
\hat{H} &=& -J\hbar(\hat{a}_{l\uparrow}^\dagger \hat{a}_{r\uparrow} + \hat{a}_{l\uparrow}^\dagger \hat{a}_{l\uparrow} + 
\hat{a}_{r\downarrow}^\dagger \hat{a}_{r\downarrow}+\hat{a}_{r\downarrow}^\dagger \hat{a}_{l\downarrow})\nonumber\\
&+&\frac{U\hbar}{2}(\hat{a}_{l\uparrow}^\dagger \hat{a}_{l\downarrow} \hat{a}_{l\downarrow}^\dagger \hat{a}_{l\uparrow} + \hat{a}_{r\uparrow}^\dagger \hat{a}_{r\downarrow} \hat{a}_{r\downarrow}^\dagger \hat{a}_{r\uparrow})\nonumber\\
\end{eqnarray}
Now, we consider our trial wave function as a linear superposition of the Fock states as follows,
\begin{eqnarray}
|\psi(t)\rangle &=& \frac{c_1(t)}{\sqrt{2}}(|\uparrow,\downarrow\rangle + |\downarrow,\uparrow\rangle) + c_2(t)|\uparrow \downarrow,0\rangle
+ c_3(t)|0,\uparrow \downarrow\rangle
\end{eqnarray}
where the states $|\uparrow , \downarrow\rangle$ , $|\downarrow , \uparrow\rangle$ define one fermion in the left well and another in right well and $|\uparrow \downarrow , 0\rangle$ , $|0 , \uparrow \downarrow\rangle$ define both fermions are in the left well and right well respectively. Here $c_1(t)$, $c_2(t)$ and $c_3(t)$ are the probability amplitudes of finding one fermion in one well, both fermions in the left and in the right well respectively. Now, putting equation $(25)$ and $(26)$ into Schr\"odinger equation, we get
\begin{eqnarray}
i\frac{dc_1}{dt} &=& -\sqrt{2}J(c_2 + c_3) + \frac{U}{2} c_1\nonumber\\
i\frac{dc_2}{dt} &=& -\sqrt{2}J c_1 + \frac{U}{2} c_2\\
i\frac{dc_3}{dt} &=& -\sqrt{2}J c_1 + \frac{U}{2} c_3\nonumber
\end{eqnarray}
By solving equation $(27)$, we get
\begin{eqnarray}
 c_1(t) &=&  A\exp\left[-i(\frac{\bar{U}}{4} - \Omega_{\rm{F}})\tau\right] + B\exp\left[-i(\frac{\bar{U}}{4} + \Omega_{\rm{F}})\tau\right]\nonumber\\
 c_2(t) &=& c_2(0) + i\sqrt{2}\left[A\frac{\exp[-i(\frac{\bar{U}}{4} - \Omega_{\rm{F}})\tau] - 1}{-i(\frac{\bar{U}}{4}-\Omega_{\rm{F}})}
 + B\frac{\exp[-i(\frac{\bar{U}}{4} + \Omega_{\rm{F}})\tau] - 1}{-i(\frac{\bar{U}}{4}+\Omega_{\rm{F}})}\right]\nonumber\\
 c_3(t) &=& c_3(0) + i\sqrt{2}\left[A \frac{\exp[-i(\frac{\bar{U}}{4} - \Omega_{\rm{F}})\tau] - 1}{-i(\frac{\bar{U}}{4}-\Omega_{\rm{F}})}
 + B \frac{\exp[-i(\frac{\bar{U}}{4} + \Omega_{\rm{F}})\tau] - 1}{-i(\frac{U}{4}+\Omega_F)}\right]\nonumber\\
\end{eqnarray}
where $\Omega_{\rm{F}} = \sqrt{4 + (\frac{\bar{U}}{4})^2}$,
 $ A = \frac{(\Omega_{\rm{F}} - \frac{\bar{U}}{4}) c_1(0) + \sqrt{2}(c_2(0)+c_3(0))}{2 \Omega_{\rm{F}}}$ and
 $ B = \frac{(\Omega_{\rm{F}} + \frac{\bar{U}}{4}) c_1(0) - \sqrt{2}(c_2(0)+c_3(0))}{2 \Omega_{\rm{F}}}$.
Assuming initial condition $c_3(0)=1$ and $c_1(0)=c_2(0)=0$,  i.e. both fermions are initially in the right well, we get
$ A = \frac{1}{\sqrt{2} \Omega_{\rm{F}}}$ and  $ B = - \frac{1}{\sqrt{2} \Omega_{\rm{F}}}$.

\section{Results and discussions}

We first present results on averages of phase-difference operators for bosons. 
We assume that bosons are initially in the right well.
In figure 1   
we show the average of the non-unitary Carruther-Nieto  cosine phase-difference operators 
and compare it with that of our unitary cosine phase-difference operators 
for 3 different numbers of total bosons  for the   interaction strength $\bar{U} = 0.05$. Figure 2 shows 
the same as in figure 1 but for $\bar{U} = 5$. Comparing the plots in figure 1 with those in figure 2
we notice that the average of the unitary cosine phase-difference operators deviate largely from 
that of the non-unitary ones particularly in the low energy regime.  
Figures  3 and 4  display variation of the average of the non-unitary and unitary 
 sine phase-difference operators for $\bar{U} = 0.5$   $\bar{U} = 5$, respectively, for different number of bosons. 
A comparison between the figures 3 and 4 shows that for large number of bosons and at large
interaction strength, the averages of unitary and non-unitary sine phase-difference operators
are almost similar. In terms of absolute magnitude and gross dynamical features,  
the deviations of the results for non-unitary sine phase-difference operators from 
those for unitary ones seem to be not as large as in the case of  cosine phase-difference 
operators. 

\begin{figure}
\includegraphics[height=3.25in, width=\columnwidth]{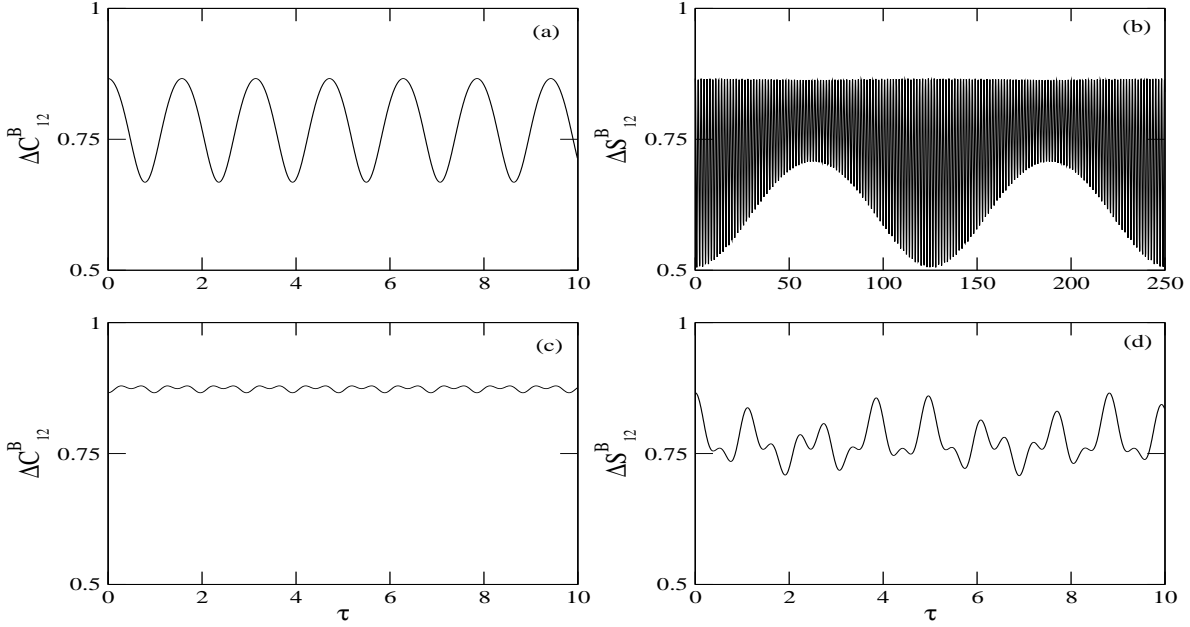}
\caption{ Fulctuations in cosine (a and c) and sine (b and d) phase-difference operators for 
a pair of bosons are plotted as a function of $\tau$ for $\bar{U} = 0.05$ (a, b) and $\bar{U} = 5$ (c and d)}. 
\label{fig9} 
\end{figure}

Using the analytical expression of equations (31) we find that for two bosons and  $\bar{U}=0$,
$\langle C^{CN}_{12}\rangle=0$ $\langle S^{CN}_{12}\rangle=\frac{1}{\sqrt{2}}\sin(2\tau)$, 
$\langle C^{B}_{12}\rangle=\frac{1}{8}(\cos(4\tau)-1)$, and  $\langle S^{B}_{12}\rangle=\frac{1}{\sqrt{2}}\sin(2\tau)$. 
For small $\bar{U}$, 
the interference between two time scale shows features like collapses and revivals in 
average quantities. For $\bar{U}\gg0$, 
the quantum mechanical average of $C^{CN}_{12}$, $S^{CN}_{12}$ and $C^{B}_{12}$,
$S^{B}_{12}$ tend to be identical. For bosons larger than 2, we find 
when $\bar{U}=0$, $\langle C^{CN}_{12}\rangle=0$ and $\langle S^{CN}_{12}\rangle\neq0$ for
any number of bosons but both terms are nonzero for $\bar{U}\neq0$.
When  $\bar{U}=0$ it is found that  $\langle C^{B}_{12}\rangle=0$ 
for odd number number of bosons  and $\langle C^{B}_{12}\rangle \ne 0$  for even number of bosons. 
For large number of bosons unitary and non-unitary phase difference operators provide almost the same results. 
In short, unitary phase-difference operators are important for low number of bosons. 

To study phase properties of fermions, we consider two-mode  fermion phase-difference  operators of equations (\ref{nfccn}) 
and (\ref{nfscn}) of a pair of two-component fermions. 
We can enumerate mainly 3 pairs of modes 
which are (a) $m \equiv l \uparrow$ and   $ m' \equiv r  \uparrow$; (b)   $m \equiv l \uparrow$ and   $ m' \equiv 
r  \downarrow$ (or  $m \equiv l \downarrow$ and   $ m' \equiv r  \uparrow$);
(c)   $m  \equiv l \downarrow$ and   $ m' \equiv r  \downarrow$; 
where $r$ and $l$ implies right and left well harmonic oscillator ground 
states.  For the mode-pair (a) there are 
two configurations  for state like  $\mid 1 0\rangle_i$ yielding 
$\sum_i \mid 1 0 \rangle_i = \mid \uparrow \downarrow, 0 \rangle +  \mid \uparrow, \downarrow \rangle$. 
 In the  case of mode-pair (b), the  number of configurations in which up spin on the left well is occupied while down
spin on the right well is empty is  only one. Therefore in the case  (b), 
we have $\sum_{i} \mid 1 0\rangle_i = \mid  \uparrow \downarrow, 0 \rangle $. We study quantum phase-difference 
for the cases (a) and (b) only. 

Figures 5 demonstrates average and fluctuations of fermionic  phase-difference operators between the two modes 
$1\uparrow$ and $2\downarrow$
for $\bar{U}=0.05$ while figure 6 exhibits the same for  $\bar{U}=5$.  From these two figures 
we notice that both the average and  fluctuation of cosine phase-difference operator 
exhibit sinusoidal behaviour as a function of time while those of sine phase-difference operators show
collapse and revivals depending on the interaction strengths. Figure 7 displays the results for average 
and fluctuation of fermionic phase-difference operators  for spin $\uparrow$ in  both the wells 
for $\bar{U}= 0.05$. Figure 8 is the counterpart of figure 7 for $\bar{U} = 5$. We infer from figure 7 and 8 
that for low $\bar{U}$, both the average and fluctuation quantities show almost  sinusoidal variation as a function 
of time. Comparing the figure 7 with  figure 5, we notice that phase fluctuations in both cosine and sine 
phase-difference operators at low 
$\bar{U}$ is larger in case of two same spin states compared to that in case of  two different spin states. To compare 
phase fluctuations of a pair of two-component fermions with those of a pair of single-component b
bosons, we display the dynamical evolution of he fluctuations 
in cosine and sine phase-difference operators for a pair of bosons in figure 9. 
Comparing figure 9 with figure 5 we notice that quantum phase fluctuation characteristics 
of pair of indistinguishable bosons in a 1D symmetric double-well potential are qualitatively different
from those of a pair of two-component fermions under similar physical condition, although 
averages of quantum phases in the two case may be qualitatively similar as can be inferred from a comparison 
between figure 3(a)  and figure 5(b). 
Note that for the 1D model considered here, 
the results for a pair of two-component fermions will be expected to be the same as that of a pair of 
two-component bosons.  However, two-component many-fermion case in higher dimensions would exhibit quantum phase properties 
different from that of two-component bosons. 

\begin{figure}
\includegraphics[height=3.25in, width=\columnwidth]{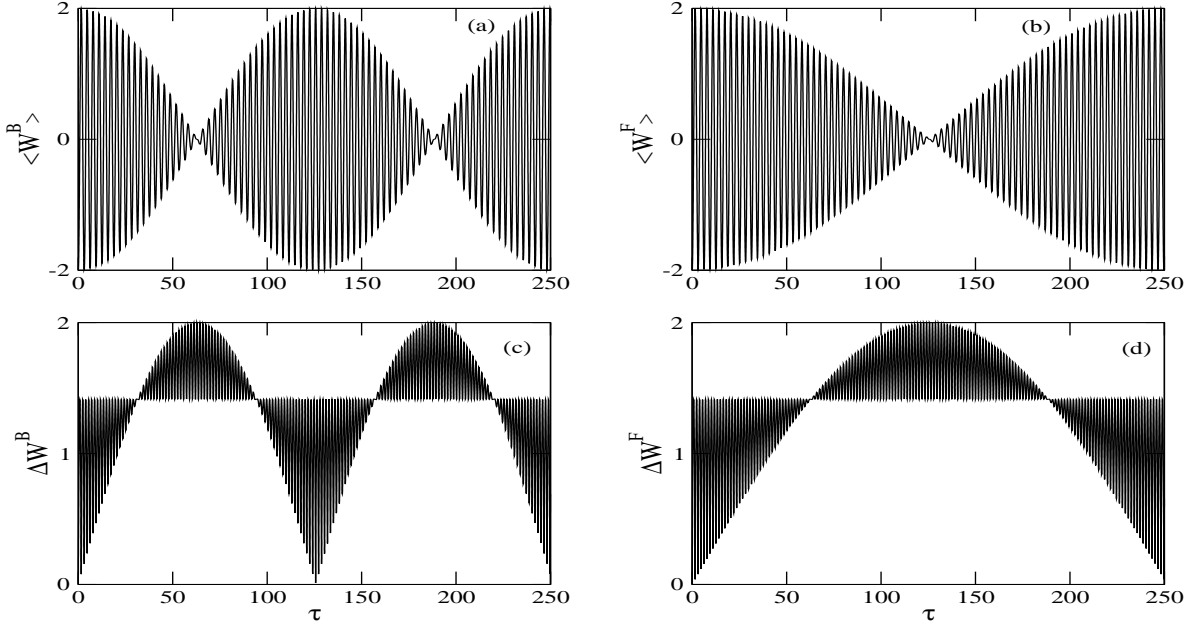}
\caption{ Shown are the plots of the average of number-difference operators for a pair of  bosons (a) and a pair of two-component 
fermions (b) for $\bar{U} = 0.5 $. The subplots (c) and (d) exhibit the fluctuations of the corresponding 
number-difference operators} 
\label{fig10} 
\end{figure}

\begin{figure}
\includegraphics[height=3.25in, width=\columnwidth]{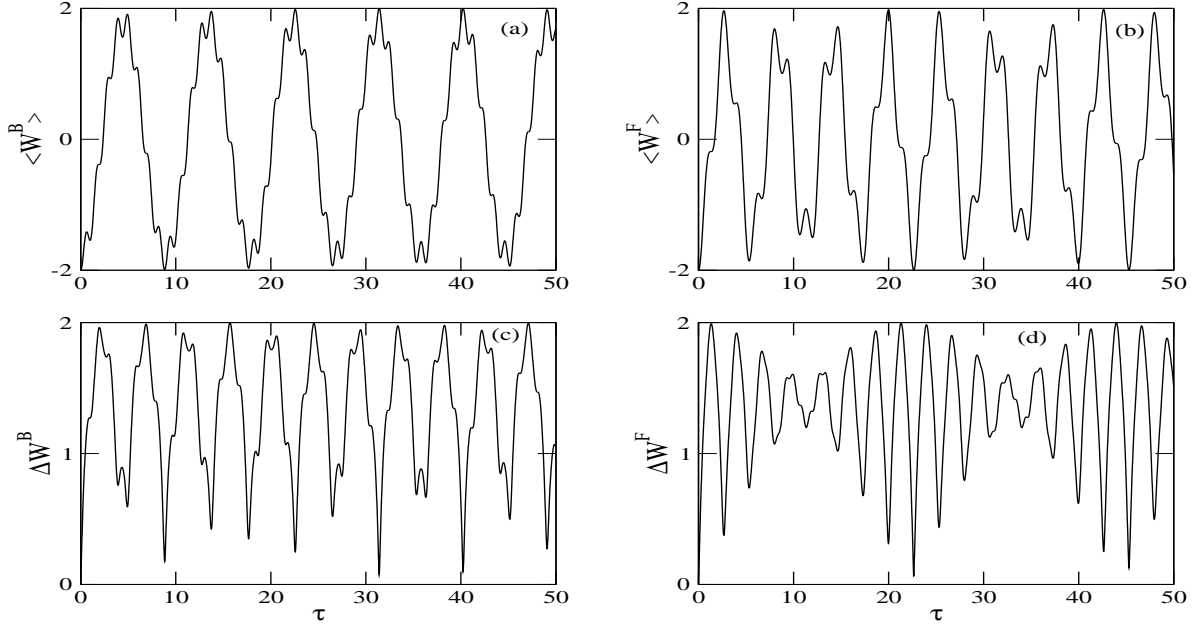}
\caption{ Same as in figure 10 but for $\bar{U} = 5$. } 
\label{fig11} 
\end{figure}

We  now present  results on average and  fluctuations  of  number-difference operators for a pair of bosons 
and a pair of two-component fermions in figures 10 and 11. 
 Note that the number-difference fluctuation 
$\Delta\hat{W}^{\rm{B}}_{12}$ is closely related to the the two-mode squeezing parameter which in turn
describes entanglement between two bosonic modes in terms of number variables. Since quantum phase-difference 
operators are canonically conjugate to number-difference operators,  
 two-mode number squeezing (reduced fluctuations in number-difference operators) would be related 
 to enhanced fluctuations in quantum phase-difference operators. In other words, 
two-mode quantum  phase fluctuations are also related to the two-mode entanglement. The entanglement between
the two modes in number variables can be quantified as the two-mode squeezing \cite{mtqo:1997:Radmore-Barnett}
or entanglement \cite{pra:2002:Burnett,pra:2002:Deb} parameter given by 
\bea
\xi_n = (\Delta(\hat n_l - \hat n_r))^2/(\lan\hat n_l\ran + \lan\hat n_r\ran)
= \frac{\left ( \Delta\hat{W}^{\rm{B}}_{12} \right )^2 }{N}
\eea
The two modes become entangled when $\xi_n$ is less than unity, or equivalently, when $\sqrt{\Delta\hat{W}^{\rm{B}}}$
becomes less than $\sqrt{N}$. 
The entanglement parameter $\xi^F_n$ for two-component fermions in terms of total fermion number fluctuation  is given by
\begin{eqnarray}
\xi^{\rm{F}}_n = 2\left[1 - \frac{2}{\Omega_{\rm{F}}^2}\sin^2(\Omega_{\rm{F}}\tau)- \frac{1}{4\Omega_{\rm{F}}^2}[\Omega_{\rm{F}}^-\cos(\Omega_{\rm{F}}^+\tau) - \Omega_{\rm{F}}^+\cos(\Omega_{\rm{F}}^-\tau)]^2\right]
\label{xif}
\end{eqnarray}
where $\Omega_{\rm{F}}^{-} = \Omega_{\rm{F}} - \bar{U}/4$, $\Omega_{\rm{F}}^{+} = \Omega_{\rm{F}} + \bar{U}/4$.
From the expression  (\ref{xif}) it is clear that  $\xi^{\rm{F}}_n$ can oscillate between 0 and 2. 
 Figure 10 (c) shows that  when the  interaction is small ($\bar U=0.05$), there are 
 time domains where 
the number-difference fluctuation remains bounded between 0 and  $\sqrt{2}$ and so the two bosons can be 
said to be entangled in those time domains. The same is also true for fermions (figure 10 (d)). Subplots 10(a) and 
10(b) exhibit that there are trapping states both for bosons and fermions when the average of number-difference 
operators vanish. We notice that near such trapping states the number-difference fluctuations are greater 
than $\sqrt{2}$ and hence when trapping occurs the two wells are not entangled in number variables. The time domains where 
entanglement occurs are far from trapping times where the average of number-difference oscillates largely. Figure 10 
shows that for large $U$ trapping does not occur and there is no time domain for entanglement.  Comparing figure 
10(d) with figure 5(d), we infer that the reduced fluctuation in fermionic  number-difference operator  with 
 enhanced  fluctuation in sine phase-difference operator or vice-versa  occur at the same time domain. This illustrates 
 number-phase uncertainty in fermionic systems in terms of unitary quantum phase operators.

\section{Conclusions }

In conclusion we have introduced Hermitian and unitary two-mode 
quantum phase-difference operators for bosons and fermions.  
Our results reveal the importance of unitary phase-operators 
in describing quantum phase properties of a few bosons or fermions.
To  the best of our knowledge, the problem of unitary quantum phase operators 
for matter waves has been addressed for the first time in this work. 
Our model studies on the comparison between non-unitary and unitary bosonic phase 
operators reveal that the results for unitary phase operators are substantially different 
from those for non-unitary ones particularly in the case of a low number of bosons. 
In our quantum phase formalism, 
the unitarity of phase operators is ensured by coupling vacuum state of one 
mode of matter waves with the highest number state of the same mode in a 
finite dimensional Fock space. In case of a Bose-Einstein condensate (BEC)  in a double-well potential,
non-unitary phase operators \cite{pra:2010:Das-Sarma}  are used in theoretical studies of quantum phases
between the two BEC's in two wells. In case of large number of particles as in a 
BEC, due to statistical effects unitary and non-unitary quantum phase operators are 
expected to yield similar results because vacuum fluctuations in case of a macroscopically 
large number of particles may not play a dominant role. However, a few bosons in a double-well 
potential 
is truly a quantum system where vacuum fluctuations can not be neglected and hence unitary 
quantum phase operators are essential for measuring quantum phases of a few-body quantum system.

Using the quantum phase operators we have studied in detail the
effects of on-site interaction on quantum phase and number fluctuation properties of  interacting bosons
and a pair two-component interacting fermions in a 1D symmetric double-well potential as an example. 
In terms of number variables, both bosonic and fermionic systems exhibit
interesting inter-well entanglement properties  
which may be a potential resource for future quantum information processing with
neutral atoms in double-well optical lattices. It would be interesting 
to explore entanglement properties of a few-body quantum system in terms of these newly introduced unitary quantum phase 
operators. 
With the first  demonstration of homo-dyne detection of a fluctuating continuous variable of 
matter waves by Gross {\it et al.} \cite{nature:2011:Gross} in 2011, it might be possible 
in near future to perform experiments on the measurement of quantum phases of matter waves in a similar
manner as in Mandel's experiments that require homo-dyne or hetero-dyne detection of weak signals.
The fermionic phase operators introduced here are applicable for a many-fermion system that  
has to be  treated within a framework of configuration-interaction or other many-body formalism which 
requires a separate study.

\section{Acknowledgment}
 Biswajit Das is thankful to the Council of Scientific \& Industrial Research (CSIR), Govt. of India, for a support.

 * Present address: Bhaba Atomic Research Centre, Mumbai 400085, INDIA.


\section*{References}
\begin{thebibliography}{150}


\bibitem{prl:2004:Ketterle}
Shin Y, Saba M, Pasquini T A, Ketterle W, Pritchard D E and Leanhardt A E 2004 {\em Phys. Rev. Lett.} {\bf 92} 050405

\bibitem{nature:2007:bloch}  F\"{o}lling S,  Trotzky S,  Cheinet P C,  Feld M, Saers R,  Widera A,  Mueller T and  Bloch I 2007 
 {\em Nature}  {\bf 448}  1029


\bibitem{prl:2007:Phillips}
Sebby-Strabley J , Brown B L, Anderlini M, Lee P J, Phillips W D and Porto J V
 2007 {\em Phys. Rev. Lett.} {\bf 98} 200405
\bibitem{prsl:1927:Dirac}
Dirac P A M 1927 {\em Proc. R. Soc. A} {\bf 114} 243
\bibitem{pl:1963:Louisell}
Louisell W H 1963 {\em Phys. Lett.} {\bf 7} 60
\bibitem{p:1964:Susskind-Glogower}
Susskind L and Glogower J 1964 {\em Physics} {\bf 1} 49-61
\bibitem{rmd:1968:Carrathers-Nieto}
Carruthers P, Nieto M M 1968 {\em Rev. Mod. Phys.} {\bf 40} 2
\bibitem{jpa:1986:Pegg-Barnett}
Barnett S M and Pegg D T 1986 {\em J. Phys. A} {\bf 19} 3849
\bibitem{prl:1991:Mandel}
Noh J W, Fougeres A and Mandel L 1991 {\em Phys. Rev. Lett.} {\bf 67} 11
\bibitem{pra:1992:Mandel}
Noh J W, Fougeres A and Mandel L 1992 {\em Phys. Rev. A} {\bf 45} 1
\bibitem{prl:1993:Mandel}
Noh J W, Fougeres A and Mandel L 1993 {\em Phys. Rev. Lett.} {\bf 71} 16
\bibitem{pra:1994:Mandel}
Noh J W, Fougeres A and Mandel L 1994 {\em Phys. Rev. A} {\bf 49} 1
\bibitem{prl:2011:bloch}
Chen Y-A,  Nascimbène, Aidelsburger S M, Atala M,  Trotzky S, Bloch I. 2011
 {\em Phys. Rev. Lett.}  {\bf 107} 210405 
\bibitem{pra:1993:Deb}
Deb B, Gangopadhyay G and Ray D S 1993 {\em Phys. Rev. A} {\bf 48} 2

\bibitem{thesis:1996:Deb} Deb B 1996 Ph. D. thesis (unpublished) Jadavpur University Kolkata, India.

\bibitem{pio:1998:Perinova}
Pe\v{r}inova V, Luk\v{s} A and Pe\v{r}ina J 1998 {\bf ``Phase in Optics''}( World Scietific Series in Contemporary Chemical Physics; vol. 15) (Singapore: World Scientific Publishing co. Pte. Ltd.)

\bibitem{pra:2010:Das-Sarma}
Zhou Q and Sarma S D 2010 {\em Phys. Rev. A} {\bf 82} 041601 (R)

\bibitem{schmelcher1} Zollner S, Meyer H-D and Schmelcher P 2008 {\em Phys. Rev. A} {\bf 78} 013621
\bibitem{schmelcher2} Zollner S, Meyer H-D and Schmelcher P 2008 {\em Phys. Rev. Lett.} {\bf 100} 040401 
\bibitem{prl:2005:Albeiz}
Albiez M {\em et al.} 2005 {\em Phys. Rev. Lett.} {\bf 95} 010402
\bibitem{pra:1997:Milburn}
Milburn G J, Corney J, Wright E M and Walls D F 1997 {\em Phys. Rev. A} {\bf 55} 4318


\bibitem{prl:2008:Kierig}
Kierig E, Schnorrberger U, Schietinger A, Tomkovic J and Oberthaler M K 2008 {\em Phys. Rev. Lett.} {\bf 100} 190405

\bibitem{prl:2005:Anker}
Anker T {\em et al.} 2005 {\em Phys. Rev. Lett.} {\bf 94} 020403

\bibitem{prl:1986:Javanainen}
Javanainen J 1986 {\em Phys. Rev. Lett.} {\bf 57} 3164

\bibitem{science:2008:Trotzky}
Trotzky S {\em et al.} 2008 {\em Science} {\bf 319} 295-299
\bibitem{pra:2008:bloch}  Barmettler P,  Rey A-M,  Demler E,  Lukin M D, Bloch I,  Gritsev V 2008
{\em Phys. Rev. A} {\bf 78} 012330 
\bibitem{ajp:2011:Foot-Shotter}
Foot C J, Shotter M D 2011 {\em Am. J. Phys} {\bf 79} 7
\bibitem{jpb:2006:Anderlini}
Anderlini M, Sebby-Strabley J, Kruse J, Porto J V and Phillips W D 2006 {\em J. Phys. B : At. Mol. Opt. Phys} {\bf 39} S199-S210
\bibitem{pra:2006:Sebby}
Sebby-Strabley J, Anderlini M, Jessen P S and Porto J V 2006 {\em Phys. Rev. A} {\bf 73} 033605
\bibitem{prl:2008:Vladimir}
Stojanovic V M, Wu C, Liu V and Sarma S D 2008 {\em Phys. Rev. Lett.} {\bf 101} 125301
\bibitem{prl:2007:Lee}
Lee P J, Anderlini M, Brown B L, Sebby-Strabley J, Phillips W D and Porto J V 2007 {\em Phys. Rev. Lett.} {\bf 99} 020402
\bibitem{nature:2007:Anderlini}
Anderlini M {\em et al.} 2007 {\em Nature} {\bf 448} 452-456
\bibitem{prb:1989:Fisher}
Fisher M P A, Weichman P B, Grinstein G and Fisher D S 1989 {\em Phys. Rev. B} {\bf 40} 546
\bibitem{epl:1993:Ramakrishnan}
Sheshadri K, Krishnamurthy H R, Pandit R and Ramakrishnan T V 1993 {\em Europhys. Lett.} {\bf 22} 257-263
\bibitem{nature:2002:Greiner}
Greiner M, Mandel O, Esslinger T, Hansch T W and Bloch I 2002 {\em Nature} {\bf 415} 39-44
\bibitem{prl:1999:Jaksch}
Jaksch D, Breigel H, Cirac J, Gardiner C and Zoller P 1999 {\em Phys. Rev. Lett.} {\bf 82} 1975
\bibitem{jpb:2011:Longhi}
Longhi S 2011 {\em J. Phys. B : At. Mol. Opt. Phys} {\bf 44} 051001
\bibitem{mtqo:1997:Radmore-Barnett}
Barnett S M and Radmore P M 1997 {\bf ``Methods in Theoretical Quantum Optics''}  Oxford University Press
\bibitem{pra:2002:Burnett}
Gasenzer T, Roberts D C and Burnett K 2002 {\em Phys. Rev. A} {\bf 65} 021605 (R)
\bibitem{pra:2002:Deb}
Deb B, Agarwal G S 2002 {\em Phys. Rev. A} {\bf 65} 063618
\bibitem{nature:2011:Gross}
Gross C {\em et al.} 2011 {\em Nature} {\bf 480} 219-223
\end{thebibliography}
\end{document}